\documentclass[preprint,12pt]{elsarticle}
\usepackage{amsmath,amssymb}
\usepackage{color,soul}
\journal{Journal of Complexity}

\newcommand{\z}{\mathbb{Z}}

\begin{document}

\begin{frontmatter}
\title{On the Complexity of \\ Generalized Discrete Logarithm Problem}

\author[First]{Cem M Unsal}
\ead{cem@terpmail.umd.edu}
\address[First]{University of Maryland, Department of Mathematics, College Park, Maryland, USA}

\author[Second]{Rasit Onur Topaloglu}
\address[Second]{IBM Corporation, Hopewell Junction, New York, USA}

\begin{keyword}
    Discrete Logarithm Problem, GDLP, Complexity, Factoring, Permutation, Benes, Network, Routing, Satisfiability, 3SAT, Post-Quantum,  Cryptography, Quantum Computing
\end{keyword}

\begin{abstract}
Generalized Discrete Logarithm Problem (GDLP) is an extension of the Discrete Logarithm Problem.
The goal is to find $x\in\mathbb{Z}_s$ such $g^x\mod s=y$ for a given $g,y\in\z_s$.
Generalized discrete logarithm is similar but instead of a single base element, uses a number of base elements which does not necessarily commute with each other.
In this paper, we prove that GDLP is NP-hard for symmetric groups.
Furthermore, we prove that GDLP remains NP-hard even when the base elements are permutations of at most 3 elements.
Lastly, we discuss the implications and possible implications of our proofs in classical and quantum complexity theory.
\end{abstract}
\end{frontmatter}
\section{Introduction}
Shor's Algorithm for prime factorization and discrete logarithm (DLP) \cite{shor1994algorithms} has been one of the central topics in quantum information since its inception.
Hardness assumptions regarding prime factorization and discrete logarithms provide the basis for the security, privacy, and authentication of some of the most common communication protocols \cite{rivest1983cryptographic,diffie2022new}.
Shor's Algorithm has demonstrated that discrete logarithm is not hard for quantum computers.
In \cite{klingler2009discrete} a team of researchers has defined a slight generalization of discrete logarithm and posed the open problem of finding a polynomial time quantum algorithm for this generalization with the hope that a similar generalization of Shor's Algorithm would solve this problem in polynomial time.
They named this problem Generalized Discrete Logarithm Problem (GDLP).
However, the time complexity of this problem was left as an open problem.

GDLP for non-abelian groups has some cryptographic applications \cite{klingler2009discrete,lanel2021cryptographic} in addition to standard well-known applications of DLP for abelian groups \cite{hankerson2006guide,diffie2022new}.
In this paper, we prove that GDLP is NP-hard and therefore is a good candidate for post-quantum era cryptography and secret sharing which is not the case for DLP due to the ease of solution with a quantum computer.
This paper is organized as follows, we first present the necessary background and definitions in section \ref{sec:back}.
Next, we discuss a sub-problem of GDLP which is distinct from DLP in section \ref{sec:clp}.
In sections \ref{sec:gdlp}-\ref{3gdlp}, we prove the NP-hardness of GDLP and some of its relevant subproblems.
In section \ref{sec:graph} we give an alternate approach to solve GDLP.
We discuss applications in post-quantum cryptography and the implications in the field of quantum information in \ref{sec:future}.

\section{Background}
\label{sec:back}
\subsection{Discrete Logarithm Problem and Its Efficient Solution}
Discrete Logarithm Problem is assumed to be hard for classical computers.
This assumption provides the theoretical basis for the security of Diffie-Hillman Key Exchange protocol \cite{diffie2022new}.
Discrete logarithm problem is the following: given $g,y\in\z_s$ and $s\in\z^*$ find $x\in\z_s$ such that
$$g^x\mod s=y.$$
Although this problem is assumed to not be solvable in polynomial time using classical computers \cite{diffie2022new}, a polynomial time algorithm exists for quantum computers \cite{shor1994algorithms}.
Shor's algorithm reduces DLP and factoring to Abelian Hidden Subgroup Problem (HSP) and 
uses a polynomial time quantum algorithm to solve Abelian HSP 
\cite{shor1994algorithms}.
However, this polynomial time reduction does not generalize to GDLP.

\subsection{Symmetric Groups and Cycle Notation}
All finite groups are isomorphic to a subgroup of a symmetric group by Cayley Theorem therefore, we can analyze GDLP in the context of symmetric groups.
Symmetric group is the set of all permutations of elements of a set of size $d$ and the standard notation for this set is $S_d$ where $d\in\z^*$.
The elements of $S_d$ are equivalent to $d\times d$ permutation matrices \cite{hungerford2012algebra}.
Just like with permutation matrices we can multiply to compose them and get another permutation. 
In this paper, we use the ``cycle notation" to denote permutations.
Cycle notation consists of the indices of elements that go to the index written to its right inside the parenthesis, separated by spaces.
For example
$$(1\quad15\quad2\quad7)$$
is the permutation where the element in index 1 goes to index 15, the element in index 15 goes to  index 2, the element in index 2 goes to index 7, and the element in index 7 goes to index 1.
As seen in this example, only the elements that change indices are written and other indices do not have to be explicitly noted.
When two or more cycles are written next to each other this means multiplication and works the same as the multiplication for permutation matrices.
We would like to note that in this paper we use zero-based indexation as our standard.
Many pure mathematicians that work on permutation groups use one-based indexation so readers should keep this in mind to avoid any confusion.
\subsection{Generalized Discrete Logarithm Problem}
Given a permutation $y$ and a vector of permutations
$$\vec{\alpha}=(\alpha_0,\alpha_1,\dots,\alpha_{q-1}),$$ 
the goal in GDLP is to find a vector of positive integers
$$\vec{x}=(x_{0,0},x_{0,1},\dots,x_{0,q-1},x_{1,0},\dots,x_{k-1,q-1})$$
for
$$\min_{\vec{x}} k$$
$$\text{s.t.}\quad\prod_{i=0}^{k-1}\alpha_0^{x_{i,0}}\alpha_1^{x_{i,1}}\dots\alpha_{q-1}^{x_{i,q-1}}=y$$
where $q,k\in \mathbb{Z}^*$ \cite{klingler2009discrete}.

When $\vec{\alpha}$ consists of a single permutation, this problem is equivalent to DLP.
However, when $\vec{\alpha}$ has more than one element, the problem takes a vastly different structure.
In this paper, we show that this problem is NP-hard.

In the decision version, the goal is to decide if $k\le \hat{k}$ for a given $\hat{k}\in \mathbb{Z}$.
For $\hat{k}=q^{O(1)}$ GDLP is in NP as we can verify the condition satisfaction for a given $\vec{x}$ of size $q^{O(1)}$ in a polynomial number of multiplications via exponentiation by squaring.
In section \ref{sec:gdlp}, we show that the decision problem is NP-hard even for $\hat{k}=1$ thereby is in the intersection of NP and NP-hard implying
that the decision version is NP-complete.
Furthermore, we show that the decision problem remains NP-complete even when $a_i$s are cycles with length restricted to be at most any integer $\ge3$ in section \ref{3gdlp}.

\subsection{Exactly-1 Positive 3-SAT}
Like the more commonly known 3-SAT problem, Exactly-1 Positive 3-SAT (Positive 1-in-3SAT) is an NP-complete problem \cite{bura2018kernel}.
We use reductions from this problem to GDLP to prove its NP-hardness.
In Positive 1-in-3SAT there are $n$ literals and $m$ clauses.
Clauses can only contain positive literals, in other words, the negation of a literal is not allowed as an input in problem instances.
The function of each clause is
$$f(u,v,w)=(u\land \neg v\land \neg w)\lor(\neg u\land v\land \neg w)\lor(\neg u\land \neg v\land w)$$
where $u,v,w$ are binary variables.
Each clause is conjugated with others by logical AND operators.
Positive 1-in-3SAT is defined by its inputs
$$u_i,v_i,w_i\in\{z_0,z_1,\dots,z_{n-1}\} \text{ for }i \in Z_n.$$
The problem is to determine if there exists $\vec{z}\in\z_2^n$ such that 
$$f(u_0,v_0,w_0)\land f(u_1,v_1,w_1)\land\dots f(u_{m-1},v_{m-1},w_{m-1})=1.$$

\begin{figure}
    \centering
    \includegraphics[width=.7\textwidth]{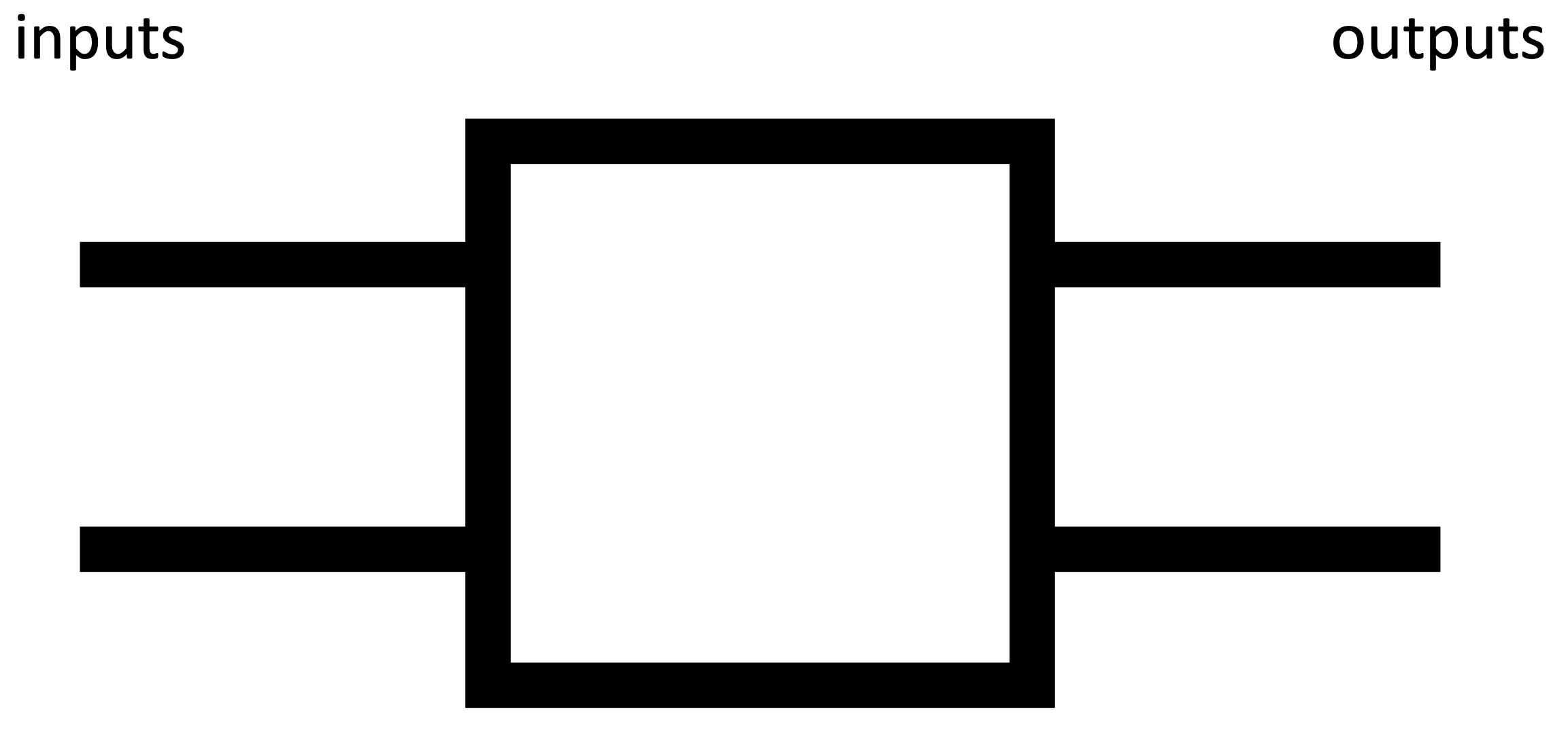}
    \caption{We use a square with 2 inputs and 2 outputs to denote a 2-by-2 switch.}
    \label{fig:switch2}
\end{figure}
\begin{figure}
    \centering
    \includegraphics[width=.7\textwidth]{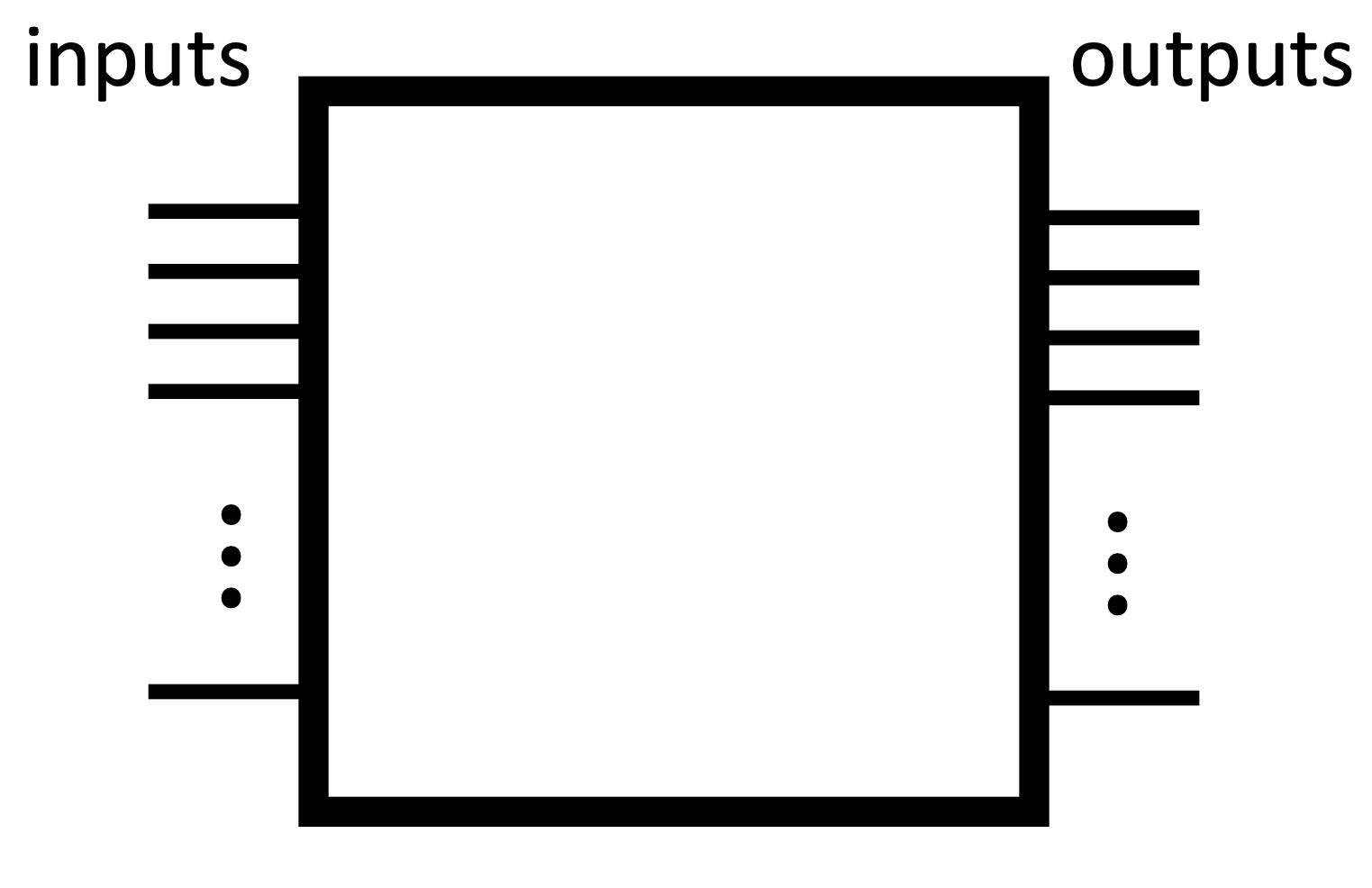}
    \caption{We use a square with h inputs and h outputs to denote an h-by-h.}
    \label{fig:switchn}
\end{figure}
\subsection{Circuit Switching and Symmetric Groups}
Circuit switching is a routing model where a dedicated channel is created between sender(s) and receiver(s) for the purposes of communication \cite{oruc1987programming}.
It is often contrasted to packet switching where instead of dedicated channels being formed packets of bits are sent to receivers via shared wires.
A permutation network is a circuit switching model where there are $h$ inputs and $h$ outputs in the circuit.
Generally, there can be anywhere between $1$ and $h$ inputs forming the sender(s).
For the purposes of this paper, we are only interested in the permutation network model that is saturated with senders, namely $h$ senders.
In the permutation network model, each sender has a distinct receiver.
We can build a permutation network with 2-by-2 switches (see Figure \ref{fig:switch2}) where the state of each switch (parallel or cross) is controlled by a single unique bit. 
If a permutation network can realize all $h!$ permutations, it is said to be an h-by-h switch (see Figure \ref{fig:switchn}).
One such network is the Complementary Benes Network that consists of $O(h\log h)$ 2-by-2 switches \cite{benevs1965mathematical}.
This is the same order of convergence as the number of bits needed to describe all $h!$ permutations.
The routing time for Complementary Benes Network is known to be $O(h \log h)$ \cite{benevs1965mathematical}.
However, larger networks exist with lower routing times.
Cellular permutation networks have $O(h^2)$ switches but only requires $O(h)$ time to route \cite{oruc1987programming}.
Considering $O(h)$ time is needed to even read the routing assignment, this is optimal.
The routing algorithm for Cellular permutation networks works by decomposing the target permutation to the successive application of permutations of size 2 via an algorithm based on the group-theoretic structure of the target permutation.
In the next section, we formulate the general circuit switching problem in terms of permutations of size 2 and the Generalized Discrete Logarithm Problem.

\section{Circuit Logarithm Problem}
\label{sec:clp}
Given a permutation circuit constructed from 2-by-2 switches, deciding if a given permutation is routable on that circuit will be referred to as Circuit Logarithm Problem (CLP).
This problem is equivalent to the decision of GDLP with $k\le1$ and bases consisting of 2 cycles.
This equivalence is computed by setting the output permutation to $y$ and the action that each switch performs to GDLP bases ($a_i$s).
In order to map the switches to the bases, we first topologically sort the switches from the inputs to the outputs.
This gives us an ordering to map to bases from left to right.
CLP is in NP but we do not know the complexity lower bound of this problem.
However, we will use it as a building block for the complexity lower bound of GDLP.

\subsection{Complementary Benes Network}
\label{sec:benes}
\begin{figure}
    \centering
    \includegraphics[width=\textwidth]{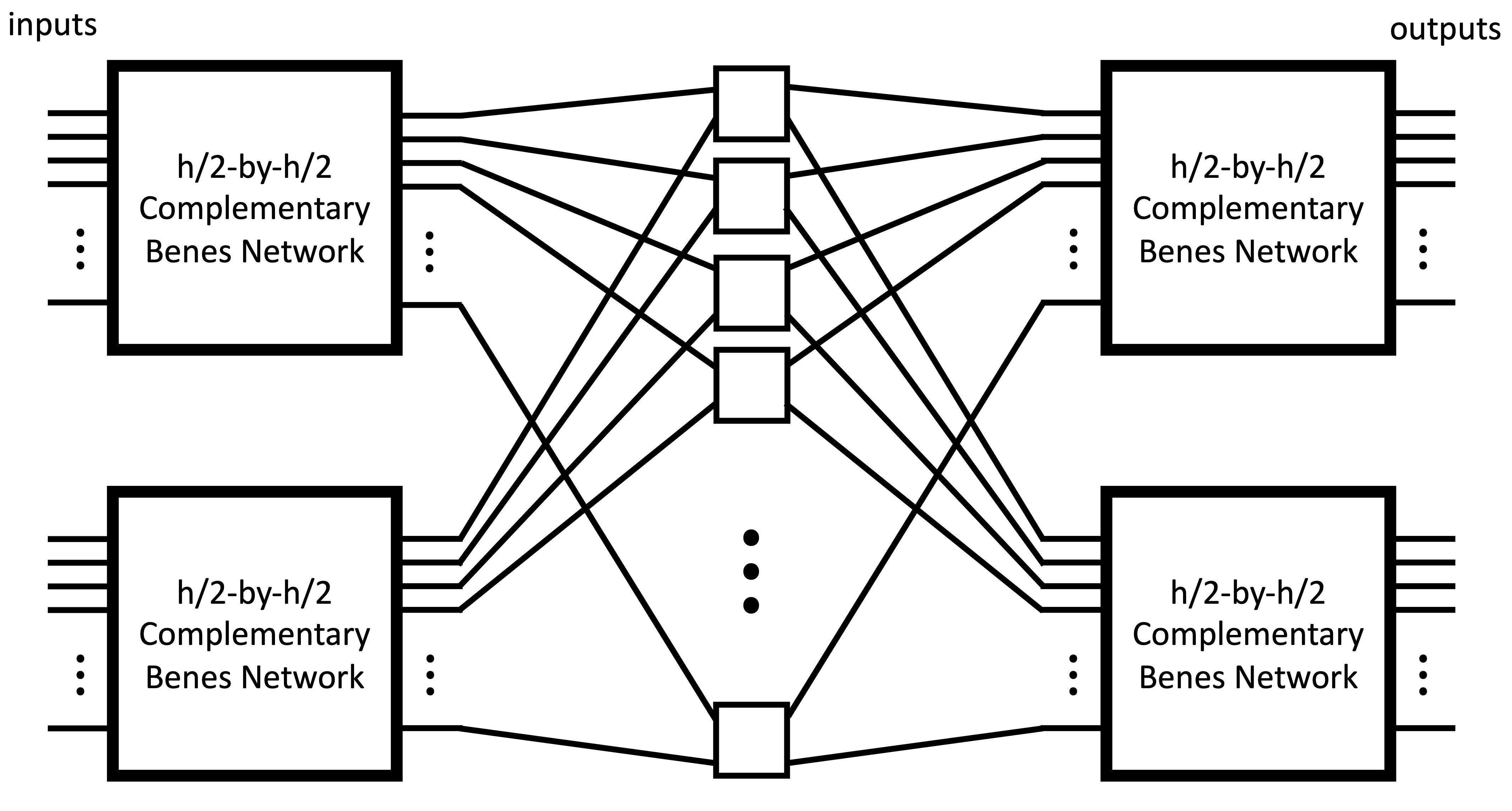}
    \caption{The recursive definition of Complementary Benes Network. If $h$ is odd, the remainder 1 gets connected from the top left to the top right without getting switched in the middle.}
    \label{fig:bensen}
\end{figure}

\begin{figure}
    \centering
    \includegraphics[width=\textwidth]{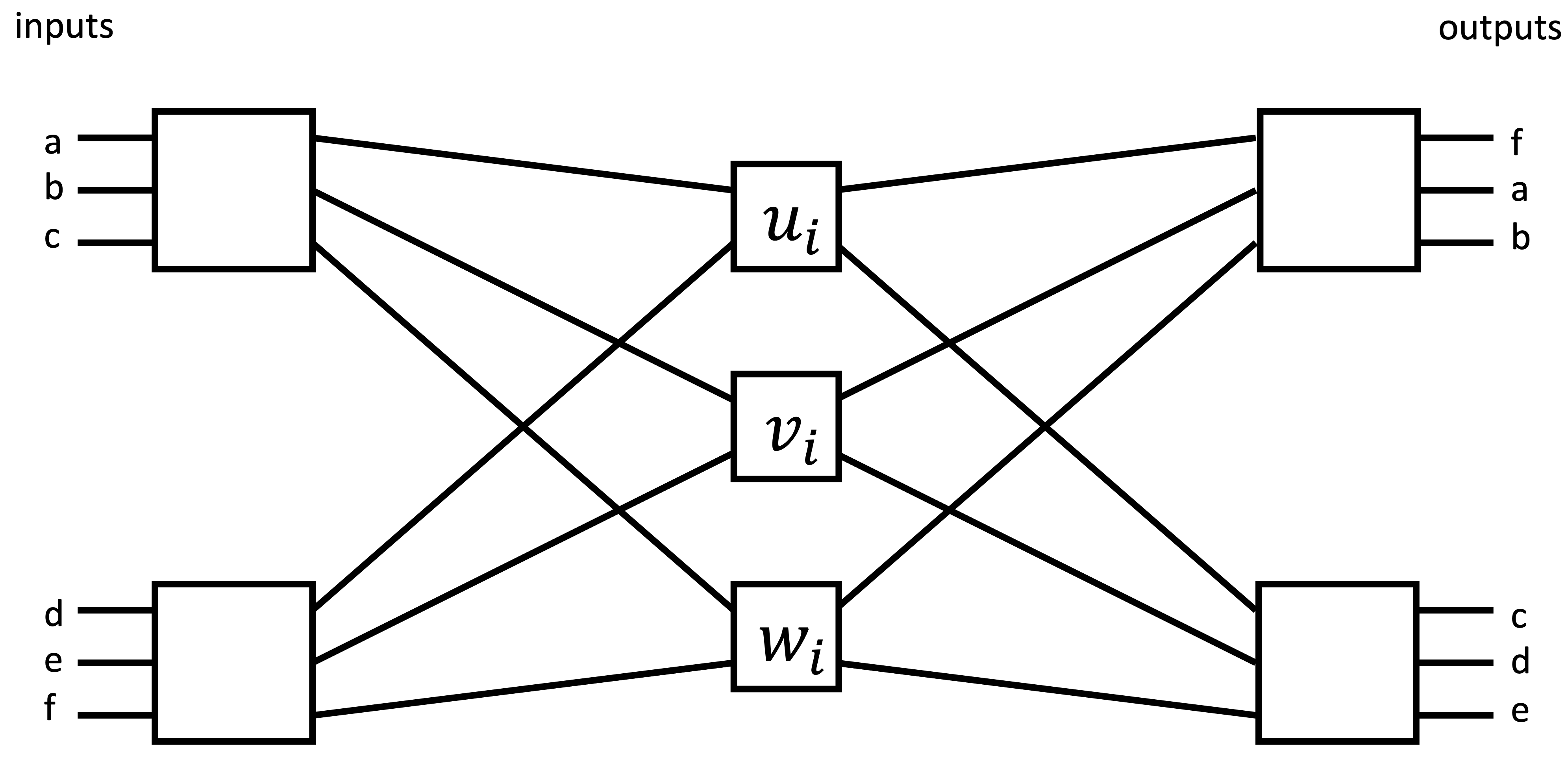}
    \caption{6-by-6 Complementary Benes Network}
    \label{fig:benes3}
\end{figure}

A Complementary Benes Network is a recursively defined h-by-h switch shown in Figure \ref{fig:bensen}.
There is an efficient algorithm to route any assignment on it \cite{benevs1965mathematical}.
The property that we care about in these networks is the fact that the number of middle 2-by-2 switches that are in the cross setting is equal to the number of elements in the bottom half of the inputs that are permuted to the top half of the outputs.
Additionally, the positions of these cross settings are arbitrary by the Slepian-Duguid theorem \cite{hui2012switching}.
We will use this fact to simulate 1 clause of Positive 1-in-3SAT.
This is done by realizing the permutation 
$$(0\,1\,2\,3\,4\,5)$$
on a 6-by-6 Complementary Benes Network (see Figure \ref{fig:benes3}).

\section{Converting Positive 1-in-3SAT to GDLP}
\label{sec:gdlp}
In this section, we construct an instance of GDLP that is $k\le1$ if and only if the given instance of Exactly-1 Positive 3-SAT is satisfiable.
This means that we can use an algorithm that determines if $k\le1$ in a given instance of GDLP to decide the satisfiability of Exactly-1 Positive 3-SAT instance.
We also show that our construction is bounded by a polynomial in the number of clauses which means that if there exists a polynomial time (in input size) algorithm for GDLP, there also exists a polynomial time algorithm for Exactly-1 Positive 3-SAT thus, all the problems in NP.
To aid with this analysis, we introduce a concept that we call ``Tethered Switches".

\subsection{Tethered Switches}
In CLP we require that each switch in the base vector $\vec{\alpha}$ is of size 2 and is independently exponentiated by the element in $\vec{x}$.
If we remove the requirement for independence of exponents, this allows us to impose global restrictions on the problem which can be used to simulate variables in Exactly-1 Positive 3-SAT that appear in more than 1 clause.
A tethered switch is a set of switches that are controlled by a single parameter.
For example, given a tether-set of switches $\{S_0,S_1,\dots S_{p-1}\}$ which switch between top wires $T_0,T_1,\dots T_{p-1}$ and bottom wires $B_0,B_1,\dots B_{p-1}$ respectively, the corresponding element in $\vec{\alpha}$ is 
$$(T_0\quad B_0)(T_1\quad B_1)\dots(T_{p-1}\quad B_{p-1}).$$
Now that we have shown how 3-by-3 switches and tethered switches can be written as entries in $\vec{\alpha}$, we will show how a circuit can be constructed by these elements which correspond to the given instance of Exactly-1 Positive 3-SAT.
\subsection{Tethered Permutation Circuit Corresponding to Exactly-1 Positive 3-SAT}
We first start with how each clause can be simulated.
As mentioned in Section \ref{sec:benes}, if exactly 1 of the top elements ends up at the bottom, Complementary Benes Network is routable if and only if exactly 1 of the switches in the middle is in the cross position.
The positions of the middle switches simulate each clause of Exactly-1 Positive 3-SAT where the top switch simulates the left literal, the middle switch simulates the center literal and the bottom switch simulates the right literal.
In order to simulate all clauses of Exactly-1 Positive 3-SAT we can replicate this circuit $m$ times.
This simulates Exactly-1 Positive 3-SAT if each literal were to be unique.
In order to simulate the case where literals might not be unique, the middle switches in each complementary Benes network corresponding to a non-unique literal will be placed in the same tether set as all the other switches corresponding to that literal.
Now we are ready to write down GDLP corresponding to an arbitrary case of Exactly-1 Positive 3-SAT with $n$ literals and $m$ clauses:
$$f(u_0,v_0,w_0)\land f(u_1,v_1,w_1)\land\dots f(u_{m-1},v_{m-1},w_{m-1})$$
$$ \text{ where } u_i,v_i,w_i\in\{z_0,z_1\dots z_{n-1}\}.$$
Using a topological sort of the corresponding bases we get a GDLP instance.
First (On the left), we start with left 3-by-3 components of Complementary Benes Networks which gives us the bases (from left to right):
$$(6i+0\quad 6i+1)$$
$$(6i+1\quad 6i+2)$$
$$(6i+0\quad 6i+1)$$
$$(6i+3\quad 6i+4)$$
$$(6i+4\quad 6i+5)$$
$$(6i+3\quad 6i+4)$$
as $i$ increases from 0 to $m-1$.
Next (in the middle), we include the bases corresponding to clause variables.
Let $u^j=\{u_i|u_i=z_i\}$, $v^j=\{v_i|v_i=z_i\}$, and $w^j=\{w_i|w_i=z_i\}$.
Then, the corresponding base elements can be written as:
$$\prod_{u_i\in u^j}(i+0 \quad i+3)\prod_{v_i\in v^j}(i+1 \quad i+4)\prod_{w_i\in w^j}(i+2 \quad i+5)$$
as $j$ increases from 0 to $m-1$.
Lastly, (in the right) the following bases are once again included to represent the right 3-by-3 components of Complementary Benes Networks:
$$(6i+0\quad 6i+1)$$
$$(6i+1\quad 6i+2)$$
$$(6i+0\quad 6i+1)$$
$$(6i+3\quad 6i+4)$$
$$(6i+4\quad 6i+5)$$
$$(6i+3\quad 6i+4)$$
as $i$ increases from 0 to $m-1$.
Finally, the target permutation $y$ that we check the satisfiability for $k=1$ is 

$$(0\,\,1\,\,2\,\,3\,\,4\,\,5)(6\,\,7\,\,8\,\,9\,\,10\,\,11)\dots(6m-6$$
$$6m-5\,\,\,\,6m-4\,\,\,\,6m-3\,\,\,\,6m-2\,\,\,\,6m-1)$$.

\section{NP-hardness of 6-GDLP}
\label{sec:6gdlp}
In the previous section, we have proved that deciding $k\le1$ for GDLP that consists of bases with 2 cycles and cycles of the form:
\begin{equation}
\alpha_t=(I_{t,0}\,\,\,I_{t,0}+3)(I_{t,1}\,\,\,I_{t,1}+3)\dots(I_{t,p-1}\,\,\,I_{t,p_t-1}+3)
\label{eq:tether}
\end{equation}
is NP-hard.
In this section, we prove that deciding $k\le1$ remains NP-hard even if the bases are at most length 6 (this will be referred as 6-GDLP for the rest of the paper).
To do this, we show how to construct an instance of 6-GDLP that is $k\le1$ if and only if the GDLP problem constructed in the previous section is also $k\le1$.
If we take the problem instance from the previous section and substitute each base of the form in \ref{eq:tether} in-place with $p_t$ bases:
$$(K_{t,p_t-1}\,\,\,S_{p-1})(I_{t,0}\,\,\,I_{t,0}+3)(K_{t,0}\,\,\,S_{t,0}),$$
$$(K_{t,0}\,\,\,S_{t,0})(I_{t,1}\,\,\,I_{t,1}+3)(K_{t,1}\,\,\,S_{t,1}),$$
$$\vdots$$
$$(K_{t,p_t-2}\,\,\,\,S_{t,p_t-2})(I_{t,p_t-1}\,\,\,\,I_{t,p_t-1}+3)(K_{t,p_t-1}\,\,\,\,S_{t,p_t-1})$$
where all $K_{i,j}$s and $S_{i,j}$s are distinct permutation elements that are $\ge6m$ and; keep the same $y$ as the previous section, the truth value of $k\le1$ is kept the same.
This is because such $y$ requires the identity output pattern on $K_{i,j}$s and $S_{i,j}$s giving 2 possible configurations for each $t$, one where nothing is flipped and one where everything is flipped simulating a tethered set entry in $\vec{\alpha}$.
This gives us $15m$ bases and permutation of $12m$ elements compared to original $m$ clauses.

\section{NP-hardness of 4-GDLP}
\label{sec:4gdlp}
In the previous section, we have proved that deciding $k\le1$ for GDLP that consists of bases with 2 cycles and cycles of the form:
\begin{equation}
\alpha_r=
    (\mathcal{A}_r\,\,\,\,\mathcal{B}_r)(\mathcal{C}_r\,\,\,\,\mathcal{D}_r)(\mathcal{E}_r\,\,\,\,\mathcal{F}_r).
    \label{eq:6}
\end{equation}
In this section, we prove that deciding $k\le1$ remains NP-hard even if the bases are at most length 4 (this will be referred to as 4-GDLP for the rest of the paper).
To do this, we show how to construct an instance of 4-GDLP that is $k\le1$ if and only if GDLP problem constructed in the previous section is also $k\le1$.
If we take the problem instance from the previous section and substitute each base of the form in (\ref{eq:6}) in-place with $4$ bases from left to right:
$$(\mathcal{A}_r\,\,\,\,\mathcal{B}_r)(L_{r,0}\,\,\,\,L_{r,1})$$
$$(\mathcal{C}_r\,\,\,\,\mathcal{D}_r)(L_{r,1}\,\,\,\,L_{r,2})$$
$$(\mathcal{E}_r\,\,\,\,\mathcal{F}_r)(L_{r,2}\,\,\,\,L_{r,3})$$
$$(L_{r,0}\,\,\,\,L_{r,1}\,\,\,\,L_{r,2}\,\,\,\,L_{r,3})$$
where $L_{r,i}$s are distinct permutation elements that are $\ge15m$ and; keep the same $y$ as the previous section, the truth value of $k\le1$ is kept the same.
This is because the output pattern for inputs $L_{r,i}$ are identity therefore the only possible outcomes of this substitution are the 2 elements generated by $(\mathcal{A}_r\,\,\,\,\mathcal{B}_r)(\mathcal{C}_r\,\,\,\,\mathcal{D}_r)(\mathcal{E}_r\,\,\,\,\mathcal{F}_r)$.
We have verified this by computing all configurations through an exhaustive search of all possible outcomes.
This gives us $24m$ bases and permutation of $24m$ elements.
\section{NP-hardness of 3-GDLP}
\label{3gdlp}
In the previous section, we have proved that deciding $k\le1$ for GDLP that consists of bases with 2 cycles and cycles of the form:
\begin{equation}
    \alpha_b=(\beta_b\quad\gamma_b)(\delta_b\quad\epsilon_b)
    \label{eq:4.1}
\end{equation}
and of the form:
\begin{equation}
    \alpha_c=(\zeta_c\quad\eta_c\quad\theta_c\quad\kappa_c).
    \label{eq:4.2}
\end{equation}
In this section, we prove that deciding $k\le1$ remains NP-hard even if the bases are at most length 3 (this will be referred to as 3-GDLP for the rest of the paper).
To do this, we show how to construct an instance of 3-GDLP that is $k\le1$ if and only if the GDLP problem constructed in the previous section is also $k\le1$.
If we take the problem instance from the previous section and substitute each base of the form in (\ref{eq:4.1}) in-place with $4$ bases from left to right:
$$(\beta_b\quad\gamma_b\quad J_{b,0})$$
$$(\beta_b\quad\delta_b\quad\epsilon_b)$$
$$(J_{b,0}\quad\beta_b\quad J_{b,1})$$
$$(\beta_b\quad J_{b,0}\quad J_{b,1})$$
and substitute each base of the form in (\ref{eq:4.2}) in-place with $3$ bases from left to right:
$$(\zeta_c\quad\eta_c\quad R_c)$$
$$(\zeta_c\quad\theta_c\quad\kappa_c)$$
$$(\theta_c\quad R_c)$$
where $J_{i,j}$s and $R_i$s are distinct permutation elements that are $\ge24m$ and; keep the same $y$ as the previous section, the truth value of $k\le1$ is kept the same.
This is because the output pattern for inputs $J_{i,j}$s and $R_i$s are identity therefore the only possible outcomes of this substitution for (\ref{eq:4.1}) are the 2 elements generated by (\ref{eq:4.1}) and the only possible outcomes of this substitution for (\ref{eq:4.2}) are the 4 elements generated by (\ref{eq:4.2}).
We have again verified this via an exhaustive search of all possible outcomes.
This gives us $45m$ base elements and a permutation of $55m$ elements.
These bases are cycles of length 2 and 3 therefore we have proved that GDLP is NP-hard even when the base elements have to be length $O(1)$ primes.

\begin{figure}
    \centering
    \includegraphics[width=.7\textwidth]{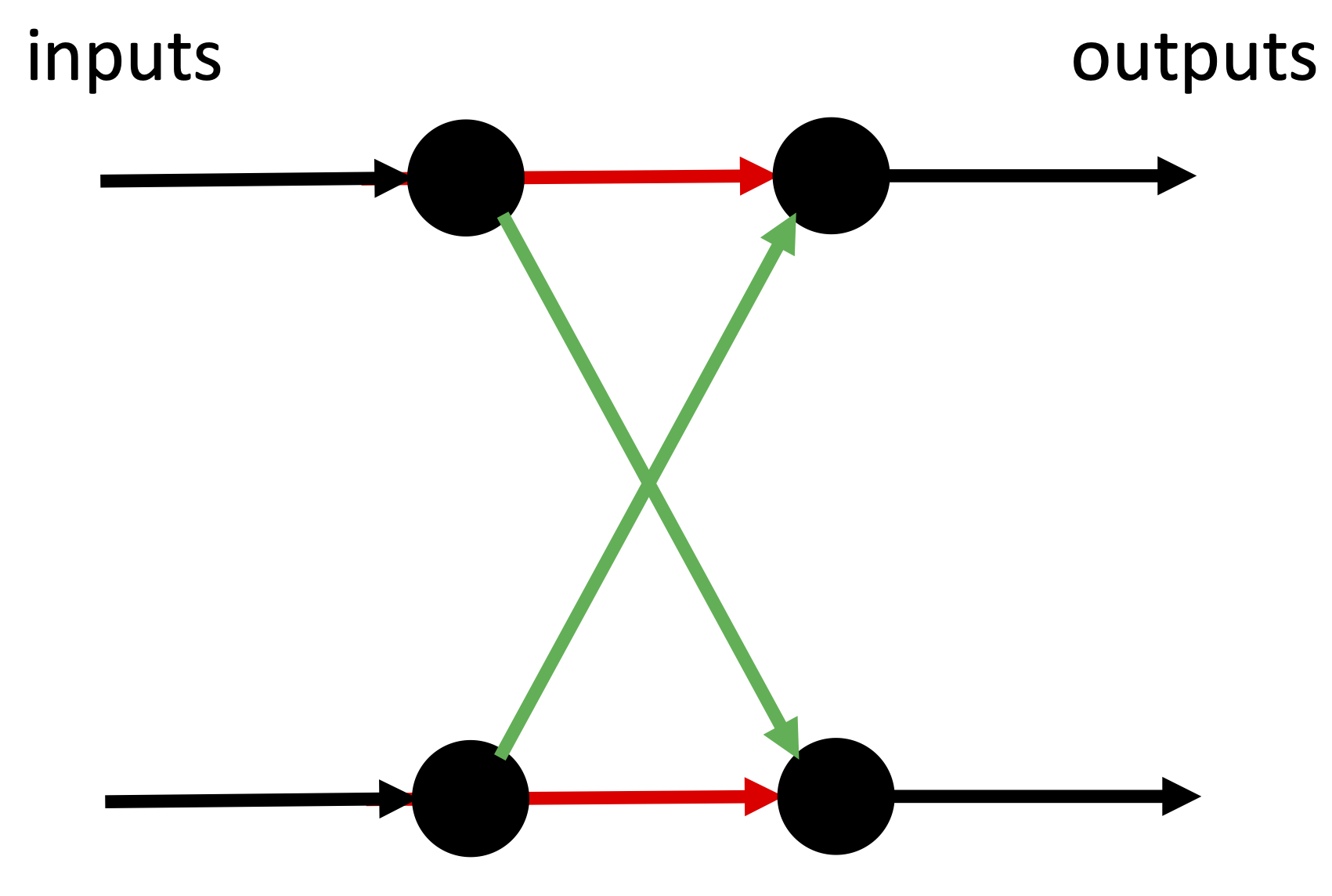}
    \caption{Routing paths corresponding to a 2 cycle}
    \label{fig:cycle2}
\end{figure}
\section{Routing on Graphs}
\label{sec:graph}
\begin{figure}
    \centering
    \includegraphics[width=.7\textwidth]{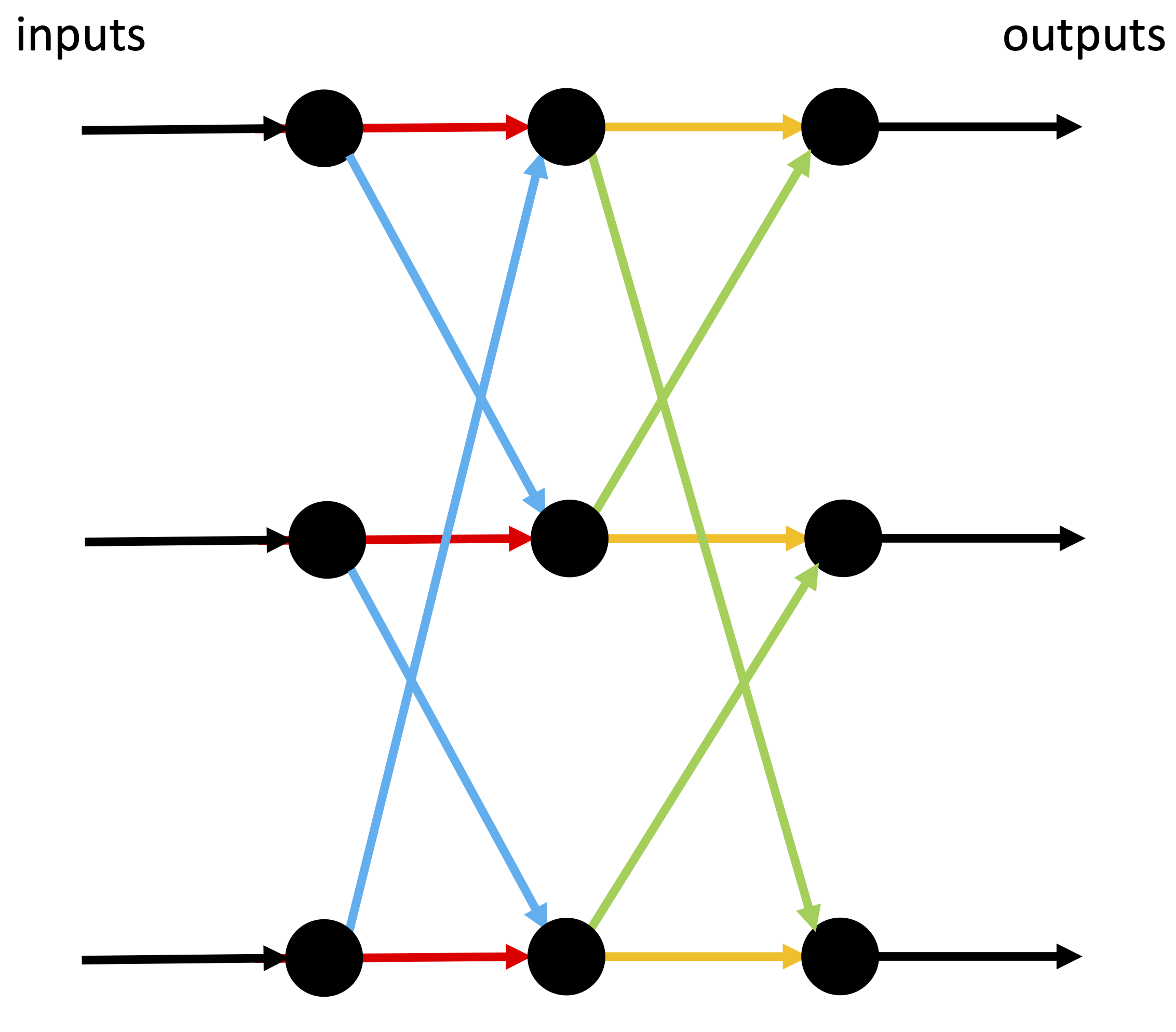}
    \caption{Routing paths corresponding to a 3 cycle}
    \label{fig:cycle3}
\end{figure}

3-GDLP can also be thought of as a type of finding non-intersecting routing paths on directed graphs.
3-GDLP consists of base elements of size 2 and size 3.
A base element of size 2 is equivalent to the 2 non-intersecting paths between inputs and outputs shown in Figure \ref{fig:cycle2} with red and blue sets of arrows.
A base element of size 3 has 3 distinct states, the identity, shift by 1, and shift by 2.
These states correspond to the non-intersecting paths between inputs and outputs shown in Figure \ref{fig:cycle3}.
Shift by 1 is the blue set of arrows followed by the yellow set.
Shift by 2 is the red set of arrows followed by the green set.
Identity is equivalent to two sets of disjoint routing paths.
The first one is red followed by yellow.
The second one is blue followed by green.
By daisy-chaining copies of these two directed graphs, it is possible to construct non-intersecting routing paths on a directed graph problem instance that is equivalent to any instance of 3-GDLP.

\section{Conclusion and Future Work}
\label{sec:future}
In this paper, we prove that GDLP for symmetric groups is NP-hard even for constant prime size cycles or GDLP with cycle size limited to any integer $\ge3$.
For potential applications, it would be essential to analyze the classical and quantum parameterized complexities of GDLP.
The relevant parameters are q, k, the number of permutation elements in all the bases, the number of bits needed to describe an instance of GDLP, and the cycle sizes of the base elements.
There are several potential applications of this fact.
First, GDLP for symmetric groups might be useful to develop homomorphic encryption schemes via Cayley Theorem.
Next, if discrete logarithm based security protocols can be extended to generalized discrete logarithm then they can be resistant to quantum cryptoanalysis if NP $\nsubseteq$ BQP.
Lastly, GDLP could be the gateway for subexponential (SUBEXP) time quantum algorithms for NP-complete problems if Shor's algorithm can be applied without pre- or post-processing exponential overhead.
In \cite{childs2003exponential} authors describe a query-efficient quantum algorithm to traverse a type of degree 2 graphs called ``welded trees".
Another way to think about that algorithm is as a routing protocol through quantum channels with "welded tree" topology.
As shown in section \ref{sec:graph}, we can formulate 3-GDLP as routing on degree 2 graphs.
So a generalization of \cite{childs2003exponential} to 3-GDLP routing graphs also has the potential to be a gateway for subexponential time quantum algorithms for NP-complete problems.
All in all, the Generalized Discrete Logarithm is an interesting structure that has potential applications  in quantum and classical computing, networking, and security.

\section{Acknowledgements}
We thank Yavuz Oruc, Andrew Childs, Elliott Lehrer, and Yusuf Alnawakhtha for valuable discussions.

\bibliographystyle{alpha}
\bibliography{gdlp.bib}

\end{document}